\documentclass[useAMS,usenatbib,usegraphicx]{mn2e}

\voffset =- .6in %for Arxiv use

   \title[Simulations of galaxy clusters]{Hydrodynamical adaptive mesh refinement simulations of turbulent flows - II. Cosmological simulations of galaxy clusters}

   \author[L. Iapichino and J. C. Niemeyer]{L. Iapichino,\thanks{E-mail:luigi@astro.uni-wuerzburg.de} and J. C. Niemeyer \\
Institut f\"ur Theoretische Physik und Astrophysik, Universit\"at W\"urzburg, Am Hubland, D-97074 W\"urzburg, Germany}

\begin{document}

\date{Accepted 2008 May 29. Received 2008 May 29; in original form 2008 January 29}

\pagerange{\pageref{firstpage}--\pageref{lastpage}} \pubyear{2008}

\maketitle

\label{firstpage}

\begin{abstract}

The development of turbulent gas flows in the intra-cluster medium
and in the core of a galaxy cluster is studied by means of
adaptive mesh refinement (AMR) cosmological simulations. A series of
six runs was performed, employing identical simulation parameters but
different criteria for triggering the mesh refinement. In particular,
two different AMR strategies were followed, 
based on the regional variability of control variables of the flow and
on the overdensity of subclumps, respectively. We show that both
approaches, albeit with different results, are useful to get an
improved resolution of the turbulent flow in the ICM. The vorticity is
used as a diagnostic for turbulence, showing that the turbulent flow
is not highly volume-filling but has a large area-covering factor,
in agreement with previous theoretical expectations. The measured
turbulent velocity in the cluster core is larger than $200\ \rmn{km\
  s^{-1}}$, and the level of turbulent pressure contribution to the
cluster hydrostatic equilibrium is increased by using the improved AMR
criteria. 

\end{abstract}

\begin{keywords}     
Hydrodynamics -- Instabilities -- Methods: numerical -- Galaxies: clusters: general -- Turbulence
\end{keywords}

\section{Introduction}
\label{intro}

The potential importance of turbulence for the physics of galaxy
clusters has been widely recognised in recent years (see the
discussion below for details and references). The precise nature of
turbulence in the intra-cluster medium 
(ICM), however, remains controversial. The relevant dimensionless parameter
for the onset of turbulence is the Reynolds number $Re$: 
\begin{equation}
\label{reynolds}
Re = \frac{LV}{\nu}\,\,,
\end{equation}
where $L$ is the integral scale of the flow instability, $V$ is the
characteristic velocity at scale $L$, and $\nu$ is the kinematic
viscosity of the fluid. For a fully turbulent flow, $Re \ga 1000$. In the 
framework of galaxy cluster physics, the uncertainty in $Re$ results
mostly from the determination of $\nu$. In the unmagnetised case, the
Braginskii formulation \citep{b65} for the viscosity is generally
used, leading to the frequently reported estimate of $Re \sim 100$
\citep[e.g.][]{snb03} in the ICM.  

This assumption does not hold in a magnetised plasma. As known from
theory \citep{s62}, the presence of magnetic fields suppresses the
transport coefficients below the unmagnetised value by a factor $f$
which depends on the tangled field structure. While highly
uncertain, it is estimated in the range $10^{-2}$ -- $1$
(\citealt{rmf05}; see also \citealt{nm01}). An effectively suppressed
($f \sim 10^{-2}$) viscosity would lead to $Re \gg 1$ both in the
cluster cores and in the hotter and less dense ICM \citep{j07}. On the
other hand, it has been claimed that a factor $f \ga 0.1$ is
consistent with the observation of filaments in the Perseus cluster
\citep{fsc03}\footnote{A weakly suppressed, or unsuppressed viscosity
  is also required for preserving the morphological stability of
  AGN-driven bubbles \citep[e.g.][]{rmf05,ss06} and for the cluster
  heating by viscous dissipation \citep{fsa03,fsc03}, though the
  numerical simulations of \citet{ss06} question the importance of
  this contribution to the cluster energy budget.}. Irrespective of a
more precise knowledge of $\nu$, it appears that the flow in the ICM
is, even with a conservative estimate, mildly turbulent.  

The turbulent nature of the flow in the intra-cluster medium (ICM)
could be directly confirmed with the help of high-resolution X-ray
spectroscopy of 
emission line broadening \citep{snb03,dvb05,bhr05,rebu07}. The
unfortunate flaw in the main instrument of the {\it Suzaku} satellite
postponed this test to the near future. Nevertheless, other
observational clues have been interpreted as evidence for the
turbulent state of the ICM, such as the analysis of pressure maps of the Coma
cluster \citep{sfm04}, the lack of resonant scattering in the $6.7\
\rmn{keV}$ He-like iron $\rmn{K_\alpha}$ line in the Perseus cluster
\citep{cfj04}, the broadening of the iron abundance profile in the
core of Perseus \citep{rcb05} and other galaxy clusters \citep{rcb06}, and
the Faraday rotation maps of the Hydra cluster \citep{ve05,ev06}. In
addition to the astrophysical problems mentioned above, an improved
knowledge of turbulence in the ICM can shed light on the amplification
of magnetic fields \citep{dbl02,brs05,ssh06}, non-thermal emission in
clusters \citep{b04}, and acceleration of cosmic rays
\citep{mjk01,mrk01,bl07}. The role of turbulence has also been
investigated as a source of heating in cooling cores, as described in
the analytical model by \citet{dc05}, where both the dissipation of
turbulent energy and the turbulent diffusion are taken into account
(cf.~also \citealt{kn03,fmw04,fsw04}). In this context, the turbulence driven by galaxy motions in the ICM has been studied e.g.~by \citet{bd89} and \citet{k07}.

From a theoretical point of view, the production of turbulence in galaxy
clusters has been ascribed to two main mechanisms. This work does not
address the outflow of active galactic nuclei (AGN), which inflate
buoyant bubbles and eventually stir the ICM, but will be focused
on turbulence produced during the hierarchical formation of
cosmic structures.  

Several numerical simulations of galaxy cluster formation show that
episodes of active merging can produce turbulence in the ICM
\citep{r98,nb99,t00,rs01,dvb05}, with typical velocities of $300$ --
$600\ \rmn{km\ s^{-1}}$ and injection scales of $300$ -- $500\
\rmn{kpc}$. In their simulations, \citet{dvb05} use a low-viscosity
version of the {\sc gadget-2} SPH implementation \citep{spri05}, which helps
to better resolve turbulent flows. \citet{ams06} showed that grid
methods are more suitable than SPH in modelling dynamical
instabilities. 

Since the properties of grid-based methods with regard
to modelling turbulence are substantially better understood than SPH, we intend
to explore their capability in the context of cosmological
simulations. These, in turn, rely strongly on adaptive mesh refinement
(AMR) in order to follow the evolution of strongly clumped
media. Using AMR for turbulent flows is a new and rapidly evolving
field \citep{knp06,knp07,sfh08} which we extend to
cluster simulations in this work.

The importance of a proper definition of the criteria for triggering
grid refinement in turbulent flows has been studied by
\citet{ias07} (hereafter paper I) by testing novel AMR criteria
designed for resolving turbulent flows, and applying them in a
simplified subcluster merger scenario. In this paper, we extend this work
to cosmological AMR simulations of galaxy clusters. In this framework,
we investigate how the AMR criteria tested in paper I can be
profitably (from a physical and computational point of view) used for
resolving the flow in the ICM. In AMR simulations of structure
formation, both the refinement of turbulent flows and the
the identification of small subclumps with a relatively low
overdensity \citep{ons05,hlf07} may be equally important; the relative
emphasis has to be determined on a case-by-case basis. Both aspects and the
related refinement strategies will be addressed in this work.  

We focus our analysis on one single galaxy cluster out of the
several objects forming in the simulated computational volume. An
extended study of turbulence should be based on average cluster
properties, sampling several objects of different masses
(cf.~\citealt{dvb05,vtc06}). Such a study is beyond the scope of this
work and is left for future analysis.  

The paper is organised as follows: in Sec.~\ref{simulations}, we
present the setup of the cosmological simulations and the set of
employed AMR criteria. We discuss some general properties of the simulated
cluster and present a performance comparison of the simulations in
Sec.~\ref{profiles}. In Sec.~\ref{properties}, the results are shown
together with comparisons between the different runs. The results are
summarised and discussed in Sec.~\ref{conclusions}.

\section{Details of the simulations}
\label{simulations}

Our work is based on the analysis and comparison of several
cosmological simulations. In this section, their setup and the
different AMR criteria are presented. The simulations were performed
using the AMR, grid-based hybrid (N-Body plus hydrodynamical) code
{\sc enzo} \citep{obb05}\footnote{{\sc enzo} homepage:
  http://lca.ucsd.edu/portal/software/enzo}.  
%The code solves the unviscid Euler equations, but some level of numerical viscosity is unavoidably introduced in our numerical scheme. 
The differences between the runs and the most significant
modifications to the original public source code are presented in
Sec.~\ref{criteria}.

\subsection{Common features}
\label{features}

We performed hydrodynamical simulations in a flat
$\Lambda$CDM background cosmology with $\Omega_\rmn{m} = 0.3$, $\Omega_\rmn{b} =
0.04$, $h = 0.7$, $\sigma_8 = 0.9$, and $n=1$. 
%, where $\Omega_{\rmn m}$ and  $\Omega_{\rmn b}$ are the fractions of critical density of matter and baryons, $h$ the Hubble parameter in units of $100\ {\rmn km\ s^{-1}\ Mpc^{-1}}$, $\sigma_8$, $n$ are the fractions of critical density of matter and baryons, .  
The simulations were started with the same initial conditions at redshift $z_{\rmn{in}} = 60$, using the \citet{eh99} transfer
function, and evolved to  $z = 0$. Cooling physics, feedback and
transport processes are neglected. An ideal equation of state was used
for the gas, with $\gamma = 5/3$. 

The simulation box had a comoving size of $128\ \rmn{Mpc}\ h^{-1}$.
It was resolved with a root grid (AMR level $l = 0$) of $128^3$ cells
and $128^3$ N-Body particles. A static child grid ($l = 1$) was nested
inside the root grid with a size of $64\ \rmn{Mpc}\ h^{-1}$,
$128^3$ cells and $128^3$ N-Body particles. The mass of each of the
particles in this grid was $9 \times 10^{9}\ M_{\sun}\ h^{-1}$. Inside
this grid, in a volume of $(38.4\ \rmn{Mpc}\ h^{-1})^3$, grid refinement from
level $l = 2$ to $l = 7$ was enabled according to the criteria
prescribed in Sec.~\ref{criteria}. The linear refinement factor $N$
was set to 2, allowing an effective
resolution of $7.8\ \rmn{kpc}\ h^{-1}$ at maximum refinement level. 

The static and dynamically refined grids were nested
around the place of formation of a galaxy cluster, identified in a
previous low-resolution, DM-only simulation using the HOP algorithm
\citep{eh98}. This cluster had a virial mass $M_\rmn{vir} = 5.8
\times 10^{14}\ M_{\sun}\ h^{-1}$ and a virial radius $R_\rmn{vir} =
1.35\ \rmn{Mpc}\  h^{-1}$ to within 1.3\% and 0.4\%,
respectively, for all simulations. At $z = 60$, the dynamically
refined part of the computational domain contained more than $99.5\%$
of the N-Body particles that were within a virial radius from the
cluster centre at $z = 0$.

\subsection{AMR criteria and other features}
\label{criteria}

An overview of the most relevant features of the performed
cosmological simulations is presented in Table \ref{run}. These runs
differ with respect to the AMR criteria that were used after $z =2$
(with the exception of run $F$). Test calculations using the new
criteria starting at $z = 60$ showed no significant differences.

\begin{table}
\caption{Summary of the cosmological simulations performed for this
  work. The first column reports the name of the run, the second the
  criteria used for the grid refinement from $z =2$ (discussed in the
  text). The third column contains the number of AMR grids at $z =
  0$. Besides the reference run $A$, the horizontal line divides the
  remaining simulations in two groups, as explained in the text.} 
\label{run}
\centering
  \begin{tabular}{@{}llc@{}}
  \hline
   Run    & AMR criteria  & $N_\rmn{grids}$ \\
\hline
$A$ & OD                   & 2612 \\
\hline
\hline
$B$ & OD + 1               & 3871 \\               
$C$ & OD + 2               & 3882 \\
$D$ & OD + 1 + 2           & 5358 \\
\hline
$E$ & OD, super-Lagrangian & 4100 \\
$F$ & OD, low threshold    & 5340 \\
    & from $z = 60$        &      \\
\hline

\end{tabular}
\end{table}

Until $z = 2$, all the simulations were run with the customary
refinement criteria based on the overdensity of baryons and DM. In both
criteria, a cell is refined if   
\begin{equation}
\label{threshold}
\rho_\rmn{i} > f_\rmn{i} \rho_0 \Omega_\rmn{i} N^l\,\,,
\end{equation}
where $\rho_0 = 3 H_0^2 / 8 \pi G$ is the critical density. In the
case of baryons, $\rho_\rmn{i} = \rho_\rmn{b}$ (baryon density) and
$\Omega_\rmn{i} = \Omega_\rmn{b}$; in the DM case, $\rho_\rmn{i} =
\rho_\rmn{DM}$ (DM density) and $\Omega_\rmn{i} = \Omega_\rmn{DM}$.  

The overdensity factors $f_i$ for baryons and DM are crucial for the
resolution of  cosmic structures \citep{ons05}. If they are set too
high, the AMR may fail in identifying low overdensity peaks, resulting
in a deficiency of low-mass halos. In our simulations, unless
stated differently, we set $f_\rmn{b} = f_\rmn{DM} = 4.0$. This is
the same as \citet{ons05} for DM, and smaller by a factor of 2 for
baryons. This overdensity criterion for baryons and DM is shortly
named ``OD'' in Table \ref{run}. In our reference run $A$, only this
criterion is used, whereas in the other simulations it is either
modified or used in combination with other criteria.  

In Table \ref{run}, the additional simulations are subdivided into two groups,
corresponding to two different methods for better refinement of
the turbulent flow. In the first group (runs $B$, $C$ and $D$), we use
the AMR criteria based on control variables of the flow
\citep{sfh08} and already tested in
simulations of a subcluster merger (see paper I). The
criteria implement a regional threshold for triggering the refinement,
based on the comparison of the cell value of the variable
$q(\bmath{x},t)$ with the average and the standard deviation of $q$, calculated
on a local grid patch: 
\begin{equation}
\label{local}
q(\bmath{x},t) \ge \langle q \rangle_i(t) + \alpha \lambda_i(t)\,\,,
\end{equation}
where $\lambda_i$ is the maximum between the average $\langle q
\rangle $ and the standard deviation of $q$ in the grid patch $i$, and $\alpha$
is a free parameter. Similar to paper I, we
tested the square of the vorticity $\omega^2$ ($\bomega = \nabla
\bmath{\times} \bmath{v}$) and the rate of compression (the negative
time derivative of the divergence $d = \nabla \cdot \bmath{v}$) as
alternative or combined control variables. They
are labelled as AMR criteria ``1'' and ``2'' in Table \ref{run},
respectively. Preliminary tests showed that, in cosmological
simulations, these new criteria are effective only when used together
with the overdensity criterion. In run $B$, the criteria ``OD'' and
``1'', with threshold $\alpha = 6.5$, are used. The run $C$ has been
set with criteria ``OD'' and ``2'', $\alpha = 6.0$. In run $D$ the
refinement is triggered by ``OD'' and both ``1'' and ``2'', with
$\alpha_1 = 7.2$ and  $\alpha_2 = 6.2$. 

Resolving turbulence in AMR cosmological simulations involves an
additional requirement to tracking and refining turbulent flows. Since
turbulence is driven largely by cluster mergers, particular care has to be
taken to properly refine low-mass subclusters. Numerical studies
\citep{ons05,hlf07} indicate that 
the smallest halos being captured by the code depend on the
overdensity thresholds and root grid resolution, rather than on the
number of AMR levels. The second group of simulations (runs $E$ and
$F$) is devoted to explore the problem of structure resolution, and to
compare this approach with the new AMR criteria used in runs $B$, $C$
and $D$. 

Run $E$ has a super-Lagrangian correction to the overdensity
thresholds defined in equation (\ref{threshold}), i.e.: 
\begin{equation}
\label{threshold-super}
\rho_\rmn{i} > f_\rmn{i} \rho_0 \Omega_\rmn{i} N^{l(1+\phi)}\,\,,
\end{equation}
where $\phi = -0.2$. The thresholds for refinement are lower
than in run $A$, especially for higher AMR levels
(cf.~\citealt{wa07}). 

The implementation of more effective AMR overdensity criteria from $z
= 2$, as in run $E$, could produce a spurious suppression of small
subclumps if they are not resolved from their formation at high
redshift. In order to investigate this effect on the ICM turbulent
motions, we performed the run $F$ with the criterion ``OD'' and
thresholds $f_\rmn{b} = f_\rmn{DM} = 2.0$, a factor of two smaller
than run $A$, using these AMR criteria from $z = 60$.

\section{General cluster properties and AMR performance}
\label{profiles}

\begin{figure}
  \resizebox{\hsize}{!}{\includegraphics{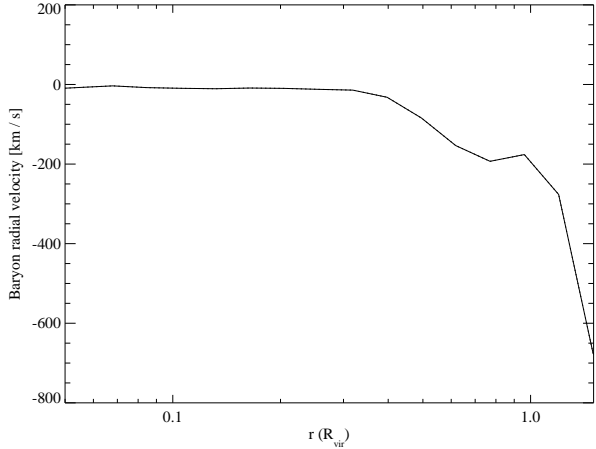}}
  \caption{Spherically averaged, mass-weighted radial velocity at $z =
    0$ in the reference run $A$.  The cluster centre, needed for this
    analysis, was determined by the HOP tool as the location of the
    local density peak, with a maximum disagreement of $52\ \rmn{kpc}\
    h^{-1}$ ($0.04\ R_\rmn{vir}$) in the performed simulations. The
    analysis tool is part of the {\sc enzo} release.} 
  \label{radial}
\end{figure}

The simulated cluster is widely relaxed, as can be inferred from its
smooth spherically averaged radial velocity profile
(Fig.~\ref{radial}) and from the time evolution of its mass accretion
which indicates a steady growth due to minor mergers and excludes
recent major merger events. In principle, this might be considered 
a poor case for studying turbulent flows in the ICM because of
the lack of violent motions driven by major merger shocks
\citep{rs01,mld05}. Nevertheless, this cluster provides a useful test
for the study of turbulent motions generated mostly by accretion of
minor subclumps, thus isolating the role of this phase of turbulence
production.  

\begin{figure}
  \resizebox{\hsize}{!}{\includegraphics{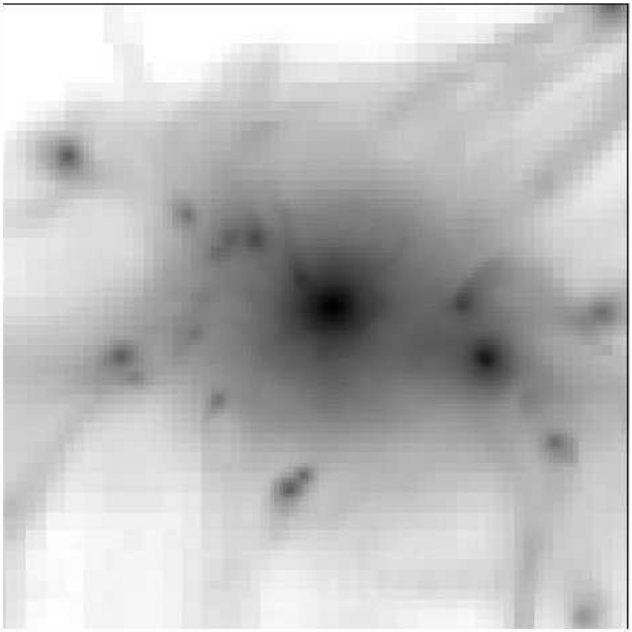}}
  \caption{Projection on the $xy$ plane of a cube with a side of $8\
    \rmn{Mpc}\ h^{-1} (\approx 6\ R_\rmn{vir})$, centred on the
    simulated galaxy cluster, and showing the baryon density at $z =
    0$ in a logarithmic gray-scale. The plot refers to run $A$, but it
    is not significantly different for other runs.} 
  \label{density-projection-a}
\end{figure}

\begin{figure*}
    \resizebox{\hsize}{!}{\includegraphics{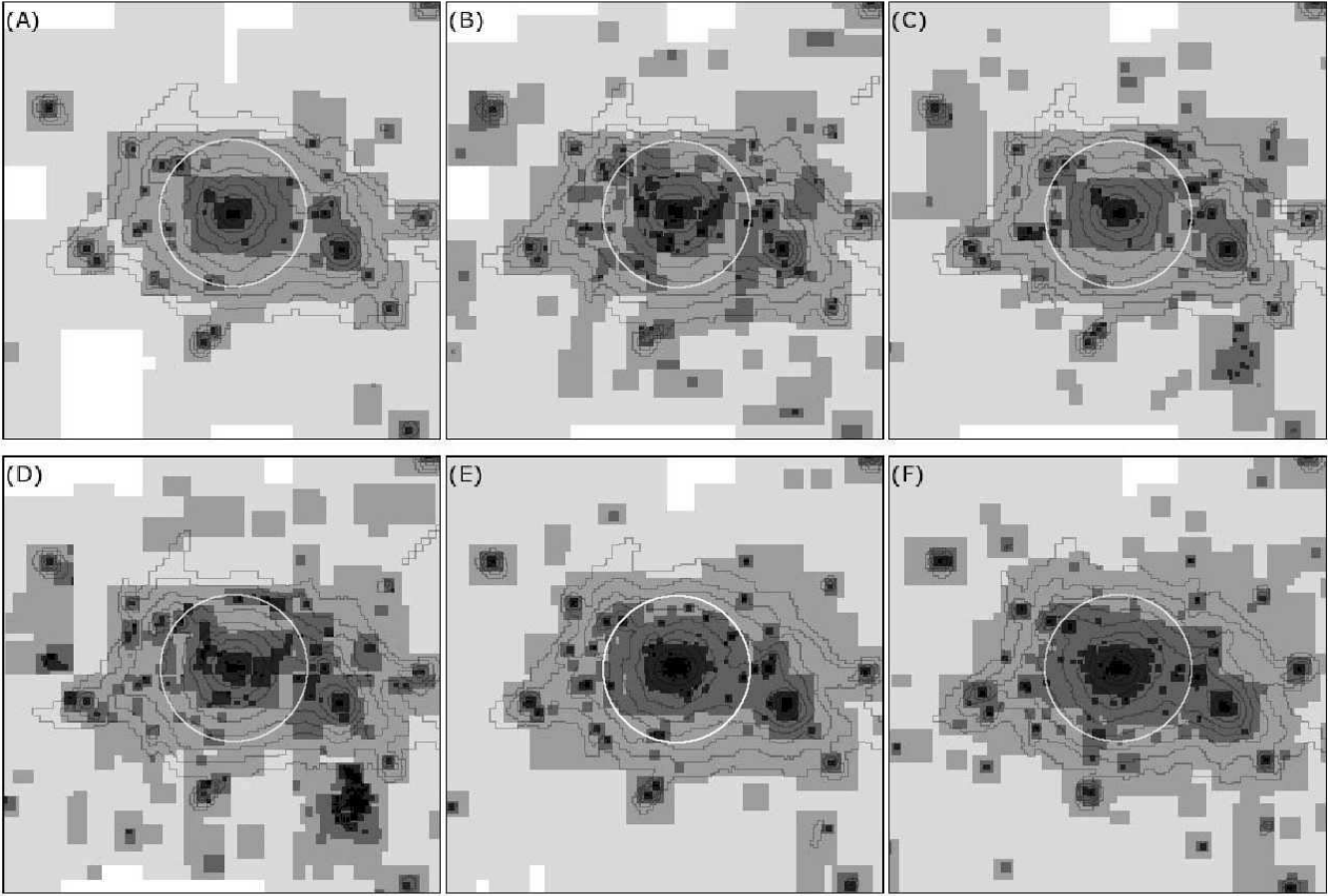}}
  \caption{Projections similar to Fig.~\ref{density-projection-a}, but
    the projected AMR level is shown. The colour scale ranges from the
    AMR level 2 (white) to 7 (black like, for example, in the inner
    regions of the cluster). Baryon density contours are superimposed
    in gray. The radius of the white circles is $r = R_\rmn{vir}$. The six panels refer to the performed runs, as indicated
    by the letter at the upper left corners.} 
\label{projections}
\end{figure*}

\begin{figure}
  \resizebox{\hsize}{!}{\includegraphics{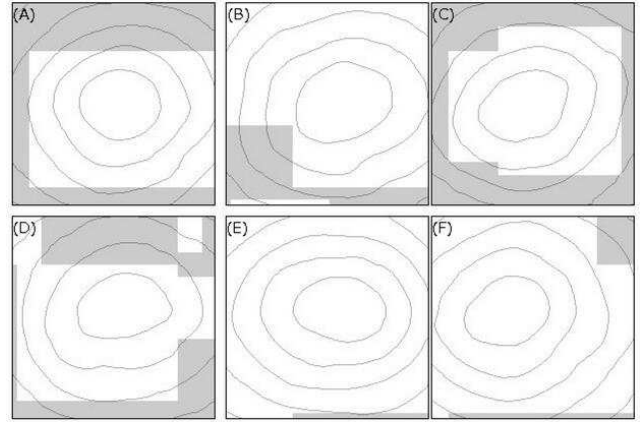}}
  \caption{Projections similar to Fig.~\ref{projections}, but showing  the projected AMR level in a zone with a side of $0.2\ R_\rmn{vir}$, centred at the cluster core. The colour white corresponds to the AMR level $l = 7$, and gray to $l = 6$. Baryon density contours are superimposed in gray. The six panels refer to the performed runs, as indicated by the letter at the upper left corners.} 
  \label{projections-core}
\end{figure}

The morphology of the simulated cluster and its outskirts is shown in
Fig.~\ref{density-projection-a}. Several accreting subclusters and the
related gas stripping are visible, together with some filaments
(e.g.~in the upper right corner). The overall structure of the cluster
does not change drastically in the performed simulations. However,
interesting differences can be observed in the projected AMR
structure (Fig.~\ref{projections})\footnote{According to the
  AMR implementation of the {\sc enzo} code, every grid patch of the same
  resolution level is handled as a single ``AMR grid''. The total
  number of such grids in the computational domain is reported in
  Table \ref{run}, third column.}.  
The AMR grid follows the density (baryons and DM) distribution in run
$A$ as well as in $E$ and $F$, where the effect of a lower AMR threshold
results in a richer grid structure. As expected, the runs performed
with the AMR criteria based on the regional variability of the control
variables of the flow show that, in addition to the refinement on the
density peaks, the grid structure is finer also in some locations
which are correlated to the local velocity fluctuations and therefore
to turbulent flows, rather than with the mass distribution. Since the analysis of the results will be largely focused on the cluster core (Sec.~\ref{core}), Fig.~\ref{projections-core} shows the projection of the AMR levels at the cluster centre at $z = 0$. This part of the computational domain is generally highly resolved, though the grid coverage changes in the different runs.

From the point of view of  grid complexity (and, consequently, of the
required computational resources), our simulations form a rather homogeneous sample. In
particular, the AMR thresholds of the simulations $B$--$F$ are tuned
in a way to produce a number of AMR grids which do not greatly exceed $2
\times N_\rmn{grids}(A)$ (cf.~Table \ref{run}).  

\begin{figure}
  \resizebox{\hsize}{!}{\includegraphics{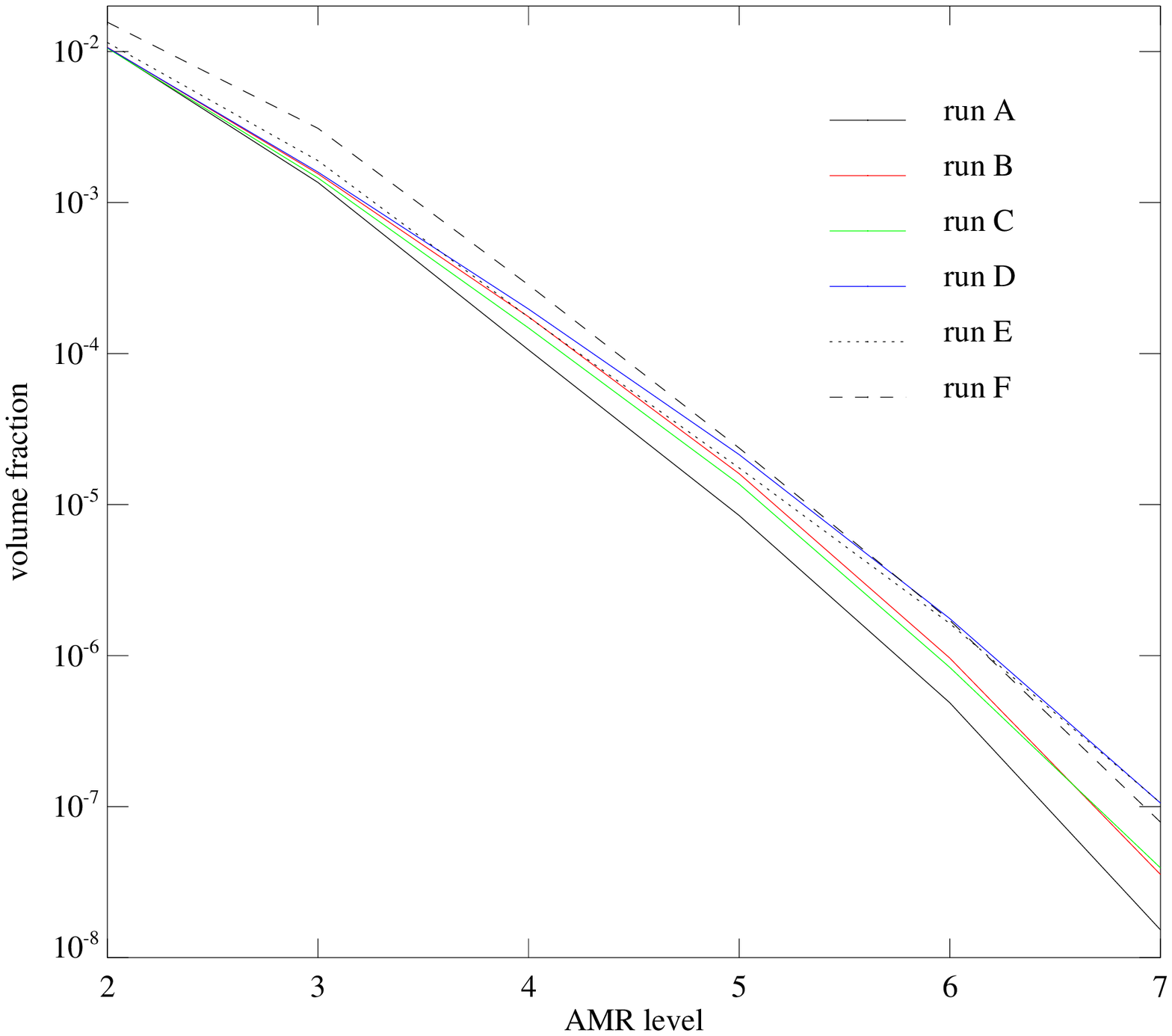}}
  \caption{Occupation fraction of the AMR levels at $z = 0$. The
    different lines refer to different simulations, according to the
    legend. The fraction is normalised to the whole computational
    domain. By construction of the setup, the occupation of level 1 is
    0.125, and AMR from level 2 is allowed in a maximum volume of
    0.027.} 
  \label{levels}
\end{figure}

As an indicator for the efficiency of AMR, the volume occupation
fraction as a function of the AMR level is shown in
Fig.~\ref{levels}. The covering differences between the AMR criteria
are more significant at high refinement levels (up to one order of
magnitude at level 7). Regarding the efficiency of resolutions at higher
levels, the simulations can be roughly grouped in three sets: the very
resolved runs $D$, $E$ and $F$, the intermediate resolved $B$ and $C$,
and the least resolved reference run $A$.  

\section{Properties of the turbulent flow in the ICM}
\label{properties}

\subsection{Resolution of subcluster mergers}
\label{mergers}

Before focusing on the features of the turbulent flow we present
a qualitative account of the effectiveness of the new AMR criteria in
refining substructures in the ICM. This analysis is directly relevant
since we are concerned with turbulence driven by infalling subclumps. 

\begin{figure*}
    \resizebox{\hsize}{!}{\includegraphics{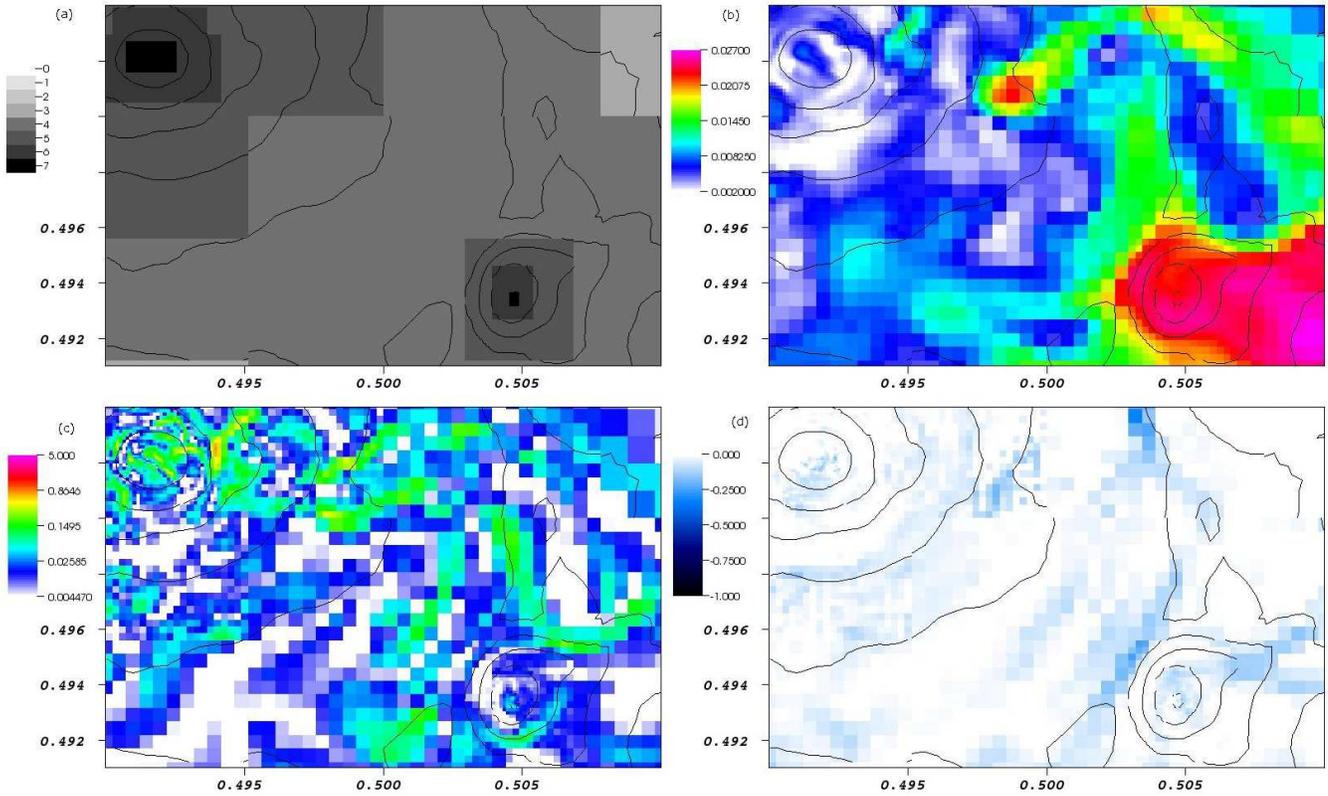}}
  \caption{Slices of the $xz$ plane for  run $A$ at $z = 0$,
    showing a detail of the simulated cluster and its surroundings. The
    coordinates are given in code units ($1 = 128\ \rmn{Mpc}\ h^{-1}$;
    $R_\rmn{vir} \sim 0.01$). In all plots, baryon density contours
    are superimposed in black. {\it (a)}: AMR levels. {\it (b)}: gas
    velocity (corrected by the centre-of-mass velocity), in code units
    ($200\ \rmn{km\ s^{-1}} \approx 0.003$). {\it (c)}: Square of the
    vorticity modulus, in code units (cf.~Sec.~\ref{filling}). {\it
      (d)}: divergence of the flow, in code units.} 
\label{merger-a}
\end{figure*}

\begin{figure*}
    \resizebox{\hsize}{!}{\includegraphics{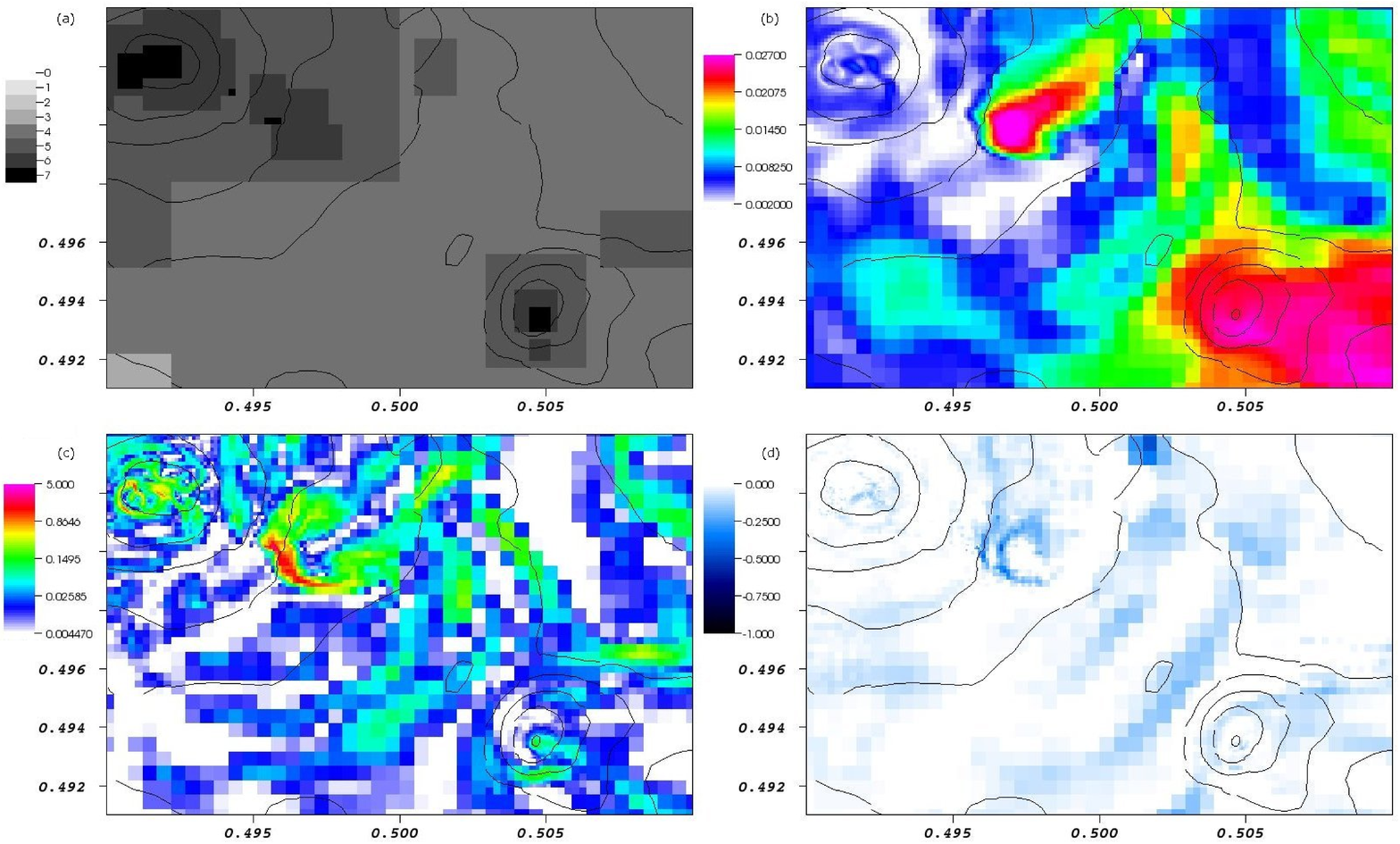}}
  \caption{Same as Fig.~\ref{merger-a}, but for the run $B$.}
\label{merger-b}
\end{figure*}

For brevity, we limit most of this comparison to the runs $A$ and
$B$. Figures \ref{merger-a} and \ref{merger-b} show some slices of the
simulated cluster and its surroundings at $z = 0$. We verified that
the results do not depend on the choice of the slice plane. From the
velocity slice (upper right panel) one can easily identify an
infalling subcluster with a mass of about $2 \times 10^{13}\ M_{\sun}\
h^{-1}$, at coordinates (0.505; 0.494), located just outside the
virial radius, and a smaller (mass of the order of some $10^{12}\
M_{\sun}\ h^{-1}$) subclump at (0.498; 0.502) moving in the ICM. The
projection of the motion of the first subcluster points roughly
towards the cluster centre, while the second one moves to the left,
parallel to the horizontal axis.  

In Fig.~\ref{merger-a}, one can observe that the refinement level in
the region of this subclump is $l = 5$. Its signature in the vorticity
plot is not very prominent. In the divergence slice, only negative
values, i.e.~converging flows, are shown, thus some of the visible
structures are shocks. For example, a distinct feature (bow shock) is
visible ahead of the larger subcluster, but it is less clear for the
smaller one.  

The application of the new AMR criteria affects the smaller
merger more strongly because it is not highly refined in run $A$. Conversely, in
run $B$ (Fig.~\ref{merger-b}) one can see that the front of the
subclump is refined to $l = 6$. The change is dramatic in the
vorticity slice, where one can identify a large increase at the merger
front and two parallel structures, likely to be caused by the lateral
shearing flow. Also the signature in the divergence plot is clear and
tracks the bow shock driven by the subclump. 

The larger subcluster, which is relatively well refined in the
reference run, does not profit further from the application of the new
AMR criteria. However, the lateral shearing flow is clearly visible
and in run $B$ a grid patch is added at (0.510, 0.496) because of
the local peak in vorticity.  

\begin{figure}
  \resizebox{\hsize}{!}{\includegraphics{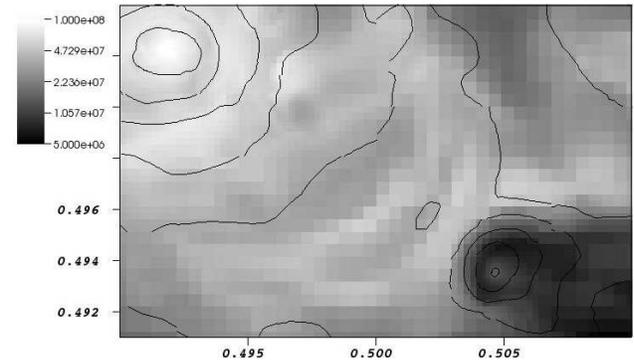}}
  \caption{Slice of the $xz$ plane, at $z = 0$, for run $B$ (same as
    the panels of Fig.~\ref{merger-b}), showing the gas temperature in
    $\rmn{K}$, as indicated by the gray-scale. Baryon density contours
    are superimposed in black.} 
  \label{temperature}
\end{figure}

It is difficult to compare the evolution of these
mergers with simulations of moving subclusters in a simplified setup
(paper I and references therein). Even if the level of effective
resolution is formally similar, the subclusters presented here have
less details than in paper I because of their smaller size, and
the background flow is far from homogeneous, in contrast with the
artificial merging setup. Nevertheless, some gross features of the
smaller subclump resemble the simplified setup, like the side
distribution of the vorticity. With the use of the new AMR criteria,
it is also easier to visualise the mergers in the temperature
slice of run $B$ (Fig.~\ref{temperature}), where they show the well known cold
front morphology. 

\begin{figure*}
    \resizebox{\hsize}{!}{\includegraphics{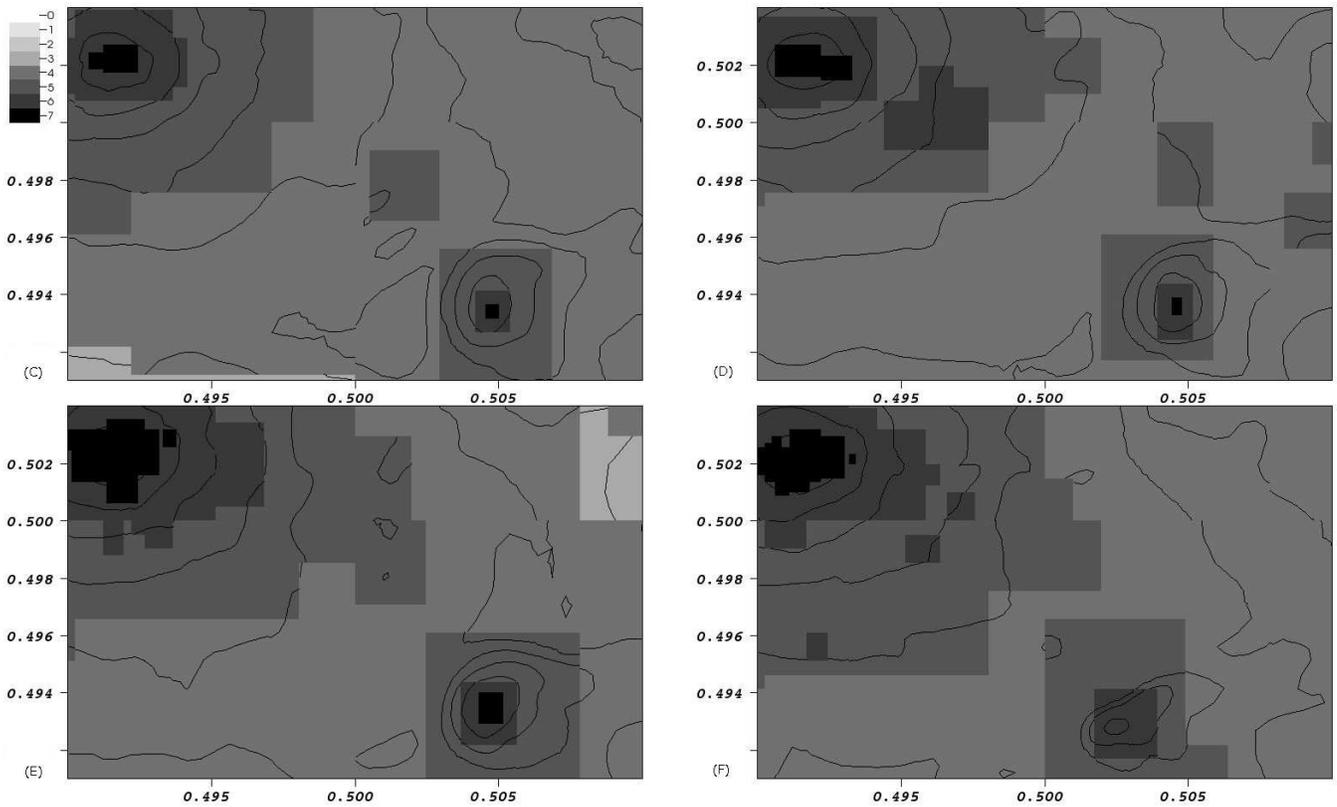}}
  \caption{Slices of the AMR levels, on the $xz$ plane, at $z = 0$,
    for the runs $C$--$F$ (same as the upper left panels of
    Figs.~\ref{merger-a} and \ref{merger-b}). The panels refer to the
    performed runs, as indicated by the letter at the lower left
    corners.} 
\label{merger-others}
\end{figure*}

For completeness, Fig.~\ref{merger-others} shows the slices of the AMR
levels for the runs $C$--$F$. The position of the smaller subclump is
resolved to $l = 6$ only in  run $D$, which implements the AMR
based on the regional variability of $\omega$ like $B$, and in run
$F$, which has a lower AMR density threshold. The run $C$ does not
refine the subclump above $l = 5$, but this is not necessarily an
indication of an insufficient performance in refining turbulent
flows. More accurate diagnostics of the resolution on velocity
fluctuations will be introduced in the next sections.  

\subsection{Volume-filling and area-covering factors}
\label{filling}

Using analytical models, \citet{ssh06} study the dynamics of turbulence generation in clusters during the infall of minor mergers. Under reasonable estimates for the merger rate and assuming simple models for the geometry of the turbulent subcluster wakes, those authors found that the volume simultaneously occupied by merger wakes (expressed by the volume-filling factor $f_V$ as a fraction of the ICM volume) is relatively small, but the projections of the wakes on the sky plane $f_S$ cover a large fraction of the cluster area.

Given a reasonable choice of the
involved parameters, assuming a viscosity suppression factor $f
\sim 0.2$ and requiring that the wake length is of the order of $2\ R_\rmn{vir}$, \citet{ssh06} report for a cluster mass of about $10^{15}\ M_{\sun}$ the indicative values
\[
f_V \sim 0.25, \qquad f_S = O(1)
\]
which depend strongly on $f$. According to those authors, a combination of a high area-covering factor and a relatively lower volume-filling factor can reconcile the observation of ordered filaments in Perseus with indications for
turbulence in the ICM of that cluster. The idea that turbulence is not completely volume-filling in the ICM is, moreover, a further motivation to use
AMR in this problem, saving computational resources
(cf.~\citealt{knp06}). 

In order to probe these features of the turbulent flow in the
performed cosmological simulations, one needs to find a suitable way
to track turbulence. Velocity fluctuations are one of the distinctive
features of turbulence, therefore quantities related to the spatial
derivatives of velocity can be used profitably for this aim.  
 In analogy with paper I, the norm of the vorticity $|\bomega|$ will be
 used to probe the velocity fluctuations at the length scale allowed
 by the spatial resolution, as a diagnostic for the resolved turbulent
 motions. The onset of eddy-like motions is closely related to the
 baroclinic generation of vorticity, expressed by taking the curl of
 both sides of the inviscid Euler equation: 
\begin{equation}
\label{vort}
\frac{\partial \bomega}{\partial t} = \nabla \bmath{\times} (\bmath{v} \bmath{\times} \bomega) - \frac{\nabla p \bmath{\times} \nabla \rho}{\rho^2}
\end{equation}
 where $p$ is the pressure of the gas. The second term on the
 right-hand side is nonzero if the two gradients are not
 parallel, i.e.~at curved shocks \citep{krc07} and at the interface of
 infalling subclumps.  

This analysis was restricted to a sphere including the innermost
$0.5\ R_\rmn{vir}$ of the cluster, at $z = 0$. As a threshold for
flagging a computational cell as ``belonging to the turbulent flow'',
we chose the mass-weighted average of the vorticity norm in run $A$
within the central sphere, $\langle|\bomega|(A)\rangle$. The
volume-filling factor $f_V$ is thus defined as the fraction of the
analysis volume where $|\bomega| > \langle|\bomega|(A)\rangle$.  
If the vorticity is
interpreted as the inverse of the eddy turnover time \citep{krc07},
the chosen threshold corresponds to a large number ($\sim 50$) of
local eddy turnovers, assuring a conservative estimate of the extent
of the turbulent regions.  

The results of this analysis are summarised in Table \ref{vorticity}. The
vorticity is reported in {\sc enzo} code units. Dimensionally, the vorticity
is expressed as $[t^{-1}]$, and the time unit in {\sc enzo} is $1/(4 \pi G
\Omega_{\rmn{m}} \rho_0 (1+z_{\rmn{in}})^3)^{1/2}$, with the meaning
of the symbols introduced above. 

\begin{table}
\caption{Results of the vorticity analysis. The first column reports
  the name of the run, the second the mass-weighted average of the
  vorticity norm $|\bomega|$ in the sphere with radius $0.5\
  R_\rmn{vir}$. The maximum value of $|\bomega|$ is listed in the third
  column, and the volume-filling factor is reported in the fourth. The
  vorticity data are expressed in code units.} 
\label{vorticity}
\centering
  \begin{tabular}{@{}lccc@{}}
  \hline
   Run & $\langle |\bomega| \rangle$ & $\max{(|\bomega|)}$ & Volume-filling  \\
          & (code units)         & (code units)        & factor          \\
\hline
$A$ & 0.211   & 1.42  & 0.233 \\  
\hline
\hline
$B$ & 0.209   & 4.04  & 0.231  \\               
$C$ & 0.215   & 2.05  & 0.243  \\  
$D$ & 0.226   & 2.75  & 0.279  \\ 
\hline
$E$ & 0.238   & 1.59  & 0.297  \\
$F$ & 0.246   & 1.49  & 0.297  \\
\hline

\end{tabular}
\end{table}

Interestingly, the volume-filling factor is not substantial and has a
value somehow comparable with the theoretical expectations of
\citet{ssh06}. Apart from the exact value of $f_V$, which
theoretically depends on many parameters and, in our simulations, on
the assumptions about the threshold of $|\bomega|$, the turbulent flow
does not appear very volume-filling ($f_V \la 0.3$ for all runs). 
 
When the results of the performed runs are compared, the features of
the vorticity are quite different in the two groups of simulations
introduced in Sec.~\ref{criteria}. In particular, in the first group
(with the exception of run $D$, to be discussed) $\langle
|\bomega| \rangle$ and $f_V$ are not much larger than for run $A$, but
$\max({|\bomega|})$ is. In the second group the opposite occurs:
$\langle |\bomega| \rangle$ and $f_V$ are larger than run $A$, while
$\max{(|\bomega|)}$ does not vary substantially. 

These results can be interpreted in terms of the difference between
runs that mostly refine on the flow (first group) and those that
refine on the substructures (second group). In the first case,
especially for run $B$ that refines explicitly only on vorticity, the
maximum resolved value of $|\bomega|$ is larger, but it does not
improve the overall efficiency in locally resolving the turbulent
flow. In the latter group, conversely, the AMR does not focus on the
turbulent flow, but the grid captures the overdense
subclumps more efficiently, which increase $\langle |\bomega| \rangle$
and $f_V$ by stirring the ICM. 
 
It is known \citep{pm01} that some spurious vorticity is generated at
the boundary of grids of different refinement levels. One could
therefore conjecture that the runs with the largest number of grids
are potentially affected by this problem. The large $f_V$ in the run
$D$ might be partly due to this issue, because it has more grids than
runs $B$ and $C$. Nevertheless, we can exclude that this effect
dominates the presented results. Evidence supporting this
statement is provided by run $E$, which has $N_\rmn{grids}$
similar (to within 5\%) to runs $B$ and $C$, but very distinct flow
properties. 

\begin{figure}
  \resizebox{\hsize}{!}{\includegraphics{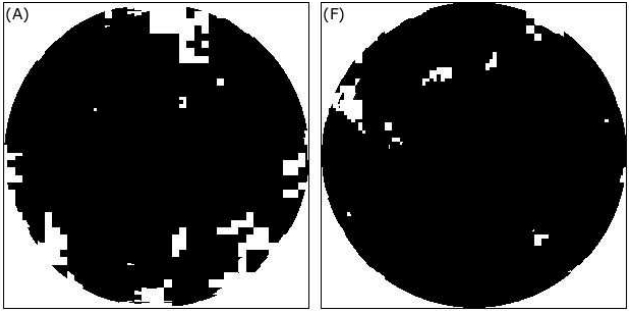}}
  \caption{Projection of the analysis sphere onto the $xy$ plane,
    showing the area-covering factor $f_S$. In the black zones,
    $|\bomega| > \langle|\bomega|(A)\rangle$ along the line of
    sight. The two panels refer to the runs $A$ (left) and $F$
    (right), as indicated by the letter at the upper left corners.} 
  \label{fs}
\end{figure}

The area-covering factor resulting from these data can be visually
inspected in Fig.~\ref{fs}, for the representative cases of runs $A$
and $F$. The turbulent flow, in agreement to \citet{ssh06}, has $f_S$
close to unity. As expected, the area-covering factor is larger in
runs with a larger $f_V$. 

\subsection{Turbulent features: the ICM and the cluster core}
\label{core}

The analysis of the geometry of the turbulent flow presented above is
complementary to the knowledge of the magnitude of the turbulent
velocity. From an operational point of view (cf.~\citealt{dvb05}), the
calculation of a root mean square (henceforth rms) velocity implies
the definition of an average reference velocity, in order to
distinguish between the bulk velocity flow and the velocity
fluctuations.    

In the case of spherically averaged radial profiles, the most natural
choice is to use the average velocity $\langle v(r) \rangle$ in the
spherical bin $r \pm \Delta r$. The mass-weighted rms baryon velocity
is then defined as 
\begin{equation}
\label{rms}
v_{\rmn{rms}}(r) = \sqrt{\frac{\sum_i m_i (v_i - \langle v(r) \rangle)^2}{\sum_i m_i}}\,\,,
\end{equation}
where $m_i$ is the mass contained in the cell $i$, and the summation
is performed over the cells belonging to the spherical shell of
amplitude $r \pm \Delta r$.  

\begin{figure}
  \resizebox{\hsize}{!}{\includegraphics{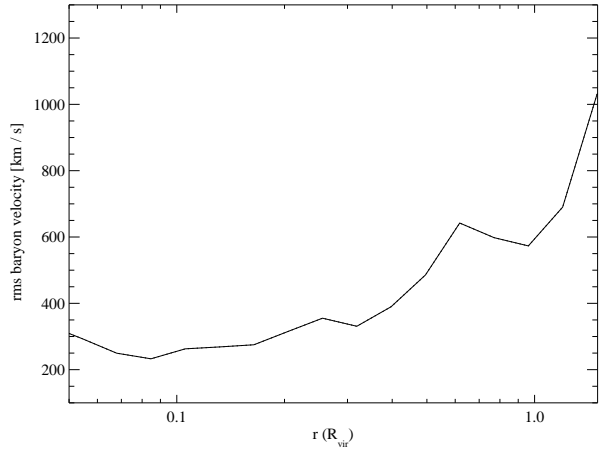}}
  \caption{Spherically averaged, mass-weighted radial profile of the
    rms baryon velocity at $z = 0$ for the reference run $A$. The
    definition of the rms velocity is provided in the text.} 
  \label{sigma_gas}
\end{figure}

The radial profile of $v_{\rmn{rms}}$ for the reference run $A$ is
shown in Fig.~\ref{sigma_gas}. The rms baryon velocity is a
fraction of the virial velocity $\sigma_{\rmn{vir}} = \sqrt{G
  M_{\rmn{vir}} / 2 R_{\rmn{vir}}} \approx 960\ \rmn{km\ s^{-1}}$.
This is similar to the results of \citet{nb99} and \citet{dvb05},
though in the latter case the quantitative comparison is more
difficult, because our cluster is outside the mass ranges which those
authors explored.  

The radial profile of $v_{\rmn{rms}}$ at $r \ga 0.1\ R_{\rmn{vir}}$
for the runs $B$--$F$ does not differ significantly from
Fig.~\ref{sigma_gas}. This result could be considered a shortcoming of
the adopted approach for the resolution of turbulent flows in the ICM,
but can be easily explained by the properties of turbulence and the
features of the performed AMR simulations. In fact, the volume filling
factor of turbulence is not very large, so any quantitative change in
$v_{\rmn{rms}}$ is likely to be washed out when averaged on a
spherical shell. Closely related to this point, the AMR thresholds imposed
to avoid an excessive number of grids particularly
affect the ICM where the volume of each spherical shell is
increasingly larger. 

In the small volume of the cluster core, where the level of refinement
is relatively higher than in the ICM, the refinement strategies are
more effective and the use of different AMR criteria introduces more
relevant changes between the simulations. We notice that also in
\citet{dvb05} (probably for reasons related to the turbulence filling
factor), the region where turbulent motions are most important is
localised in the inner $\approx 0.1\ R_{\rmn{vir}}$. In the core region of
the cluster, the analysis was not performed on radial profiles
because this tool proved not to be robust for $r \la
0.07\ R_{\rmn{vir}}$, probably because of sampling issues. In
particular, the baryon rms velocity profile in the innermost part of the
cluster is very sensitive to small displacements of the centre and to
the calculation of $\langle v(r) \rangle$. Therefore, the
mass-weighted averaged quantities within a sphere of radius $R =
0.1\ R_{\rmn{vir}}$, centred at the cluster centre, provide a more
robust diagnostic for the resolution of turbulent flows in the cluster
core. 

\begin{table*}
\caption{Mass-weighted values of some hydrodynamical quantities,
  calculated within a sphere with $R = 0.1\ R_{\rmn{vir}}$ centred at
  the cluster centre at $z = 0$. For the different runs (first
  column), the table lists the rms baryon velocity $v_{\rmn{rms}}$
  (second column), the baryon density $\rho$ (third), the temperature
  $T$ (fourth), the ratio of turbulent to total pressure
  $P_{\rmn{turb}} / P_{\rmn{tot}}$ (fifth, defined in the text) and
  the entropy $S = T / \rho^{\gamma - 1}$, with $\gamma = 5/3$
  (sixth).} 
\label{core-table}
 \centering
 \begin{minipage}{100mm}
  \begin{tabular}{@{}lccccc@{}}
  \hline
   Run &  $v_{\rmn{rms}}$ & $\rho$  & $T$ & $P_{\rmn{turb}} / P_{\rmn{tot}}$  & $S$ ($10^5\ \times$ \\
       & ($\rmn{km\ s^{-1}}$)  & ($10^{-26}\ \rmn{g\ cm^{-3}}$) & ($10^7\ \rmn{K}$)  & (\%) & code units) \\
\hline
$A$    & 211 & 1.13 & 7.85 & 1.37 & 3.11 \\ 
\hline
\hline
$B$    & 240 & 1.05 & 8.14 & 1.71 & 3.41 \\ 
$C$    & 298 & 1.04 & 7.97 & 2.69 & 3.37 \\ 
$D$    & 266 & 1.07 & 8.23 & 2.08 & 3.39 \\ 
\hline
$E$    & 239 & 1.09 & 8.04 & 1.72 & 3.26 \\ 
$F$    & 272 & 1.05 & 7.91 & 2.26 & 3.32 \\ 
\hline

\end{tabular}
\end{minipage}
\end{table*}

The results of the core analysis are summarised in Table
\ref{core-table}. For the calculation of $v_{\rmn{rms}}$, equation
(\ref{rms}) was applied, where $\langle v(r) \rangle$ is the mean
velocity in the analysis sphere. The rms velocity is the quantity that
shows the largest variation between the reference run $A$ and runs
$B$--$F$, with an increase of about 40\% for run $C$. We notice
that the highest value is predicted in a run belonging to the first group. 

Among the first group, run $C$ seems more effective than run $B$. A
possible reason is that the AMR implemented in run $B$ is not designed
to refine explicitly on shocks (cf.~Paper I), which are ubiquitous in
the cluster medium \citep{mrk00,rkh03,krc07}. Despite of the larger
number of grids, run $D$ has features which are comparable with runs
$B$ and $C$, probably because of the larger thresholds used for its
refinement  criteria.  

In the second group, run $F$ performs slightly better than run
$E$. Comparing the density thresholds for refinement in these two runs
(equations (\ref{threshold-super}) and (\ref{threshold}),
respectively, with the appropriate parameters), one can see that the
thresholds in run $E$ are smaller for $l > 5$, namely in most of the
cluster core. Therefore, the better performance of run $F$ in the core
is not due to a more intensive, local use of AMR, but is likely to be
caused by the more accurate tracking of subclumps along the whole
cluster evolution.  

Other variables are also listed in Table \ref{core-table}. For
density, temperature and entropy the variations introduced by the use
of different AMR criteria are smaller than in the case of
$v_{\rmn{rms}}$ (about 10\%, at most, for density and entropy, and 5\%
for temperature).  

The ratio of turbulent to total pressure (fifth
column in Table \ref{core-table}) is defined as: 
\begin{equation}
\label{pressure}
\frac{P_{\rmn{turb}}}{P_{\rmn{tot}}} = \frac{v_{\rmn{rms}}^2 /
  3}{k T / (\mu m_{\rmn{p}}) + v_{\rmn{rms}}^2 / 3}  \,\,,
\end{equation}
where $k$ is the
Boltzmann constant, $\mu = 0.6$ is the mean molecular weight in
a.m.u., $m_{\rmn{p}}$ is the proton mass, and $T$ is the
mass-weighted temperature (fourth column in Table \ref{core-table}).

The contribution of $P_{\rmn{turb}}$ to the total pressure in the
cluster centre is marginal, but we notice that it is increased
(in case of run $C$, almost doubled) in the new runs. The increase of
$P_{\rmn{turb}}$ is consistent with the decrease in density: the
turbulent motions introduce an additional term to the pressure
equilibrium, which is balanced by a smaller baryonic pressure and a
smaller central density.  

An increase in temperature and entropy with respect to
run $A$ are also observed. The moderate increase in entropy is similar
to what is observed in \citet{dvb05}, when the low-viscosity version
of SPH is used. In that work, the increase was ascribed either to a
better shock modelling or to a more efficient gas mixing.  Both
features can be retrieved in our AMR simulations $B$--$F$, at some
degree, so we agree with this interpretation.  

The trend in temperature is compatible with the enhanced dissipation
of kinetic to internal energy, although the increase is not closely
related with the value of $v_{\rmn{rms}}$. We verified that
the increase in the internal energy in the new runs is comparable, by
order of magnitude, to the increase of the energy dissipation $\Delta
E = (\Delta v)^3 \tau / 2 \delta$, where $\Delta v$ is the increase of
$v_{\rmn{rms}}$, $\delta$ is the effective spatial resolution and
$\tau$ is a time on the Gyr-scale. 

\begin{figure}
  \resizebox{\hsize}{!}{\includegraphics{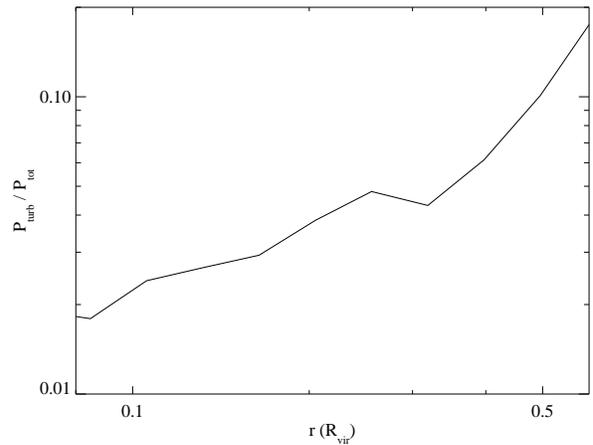}}
  \caption{Spherically averaged, mass-weighted radial profile of the
    ratio of turbulent to total pressure (defined in the text), at $z
    = 0$, for the reference run $A$.} 
  \label{pressure-ratio}
\end{figure}

Figure \ref{pressure-ratio} displays the radial profile of the
turbulent to total pressure outside the cluster core, for the run
$A$. Equation \ref{pressure} was used for the computation of $P_{\rmn{turb}}/P_{\rmn{tot}}(r)$, with $v_{\rmn{rms}}$ and $T$ averaged in the spherical shell $r \pm \Delta r$.  Interestingly, the profile increases with $r$ because of the
decreasing temperature profile and the slightly increasing velocity
dispersion profile. This behaviour suggests a growing importance
of the turbulent contribution to the total pressure, up to 10\% within
$R = 0.5\ R_{\rmn{vir}}$, where the ICM can be reasonably assumed in
hydrostatic equilibrium (HSE), as also indicated by
Fig.~\ref{radial}. Similar to Fig.~\ref{sigma_gas}, the pressure ratio
profile is not changed by the use of new AMR criteria. Nevertheless,
the magnitude of the turbulent pressure contribution suggests that a
better turbulence modelling (cf.~Sec.~\ref{conclusions}) could have
interesting outcomes in the ICM.   

Finally, we note that the DM velocity dispersion profile is basically
unchanged along the performed simulations, consistent with our methods
and  in agreement with \citet{dvb05}.

\subsubsection{Convergence test of the AMR criteria}
\label{convergence}

In order to perform a verification test on the new AMR criteria, it is important to check whether the increase of $v_{\rmn{rms}}$ with respect to value computed for the reference run $A$ is actually due to a better refinement of the turbulent flow, or to spurious correlations with other grid-related quantities like $N_\rmn{grids}$ or the grid coverage of the cluster core.

First, in the presented results we notice that there is no obvious correlation between the values of $v_{\rmn{rms}}$ (Table \ref{core-table}) and the grid coverage of the cluster core (Fig.~\ref{projections-core}), meaning that the magnitude of the rms velocity does not depend trivially on the level of core resolution. Analogously, it is true that run $A$ has less grids than the other runs, but $v_{\rmn{rms}}$ does not seem directly correlated with $N_\rmn{grids}$ (cf.~Table \ref{run}). This is clearly shown by the comparisons of run $B$ with $C$, or $D$ with $F$, which have almost the same $N_\rmn{grids}$ at $z = 0$ but different core velocity dispersion. 

\begin{figure}
  \resizebox{\hsize}{!}{\includegraphics{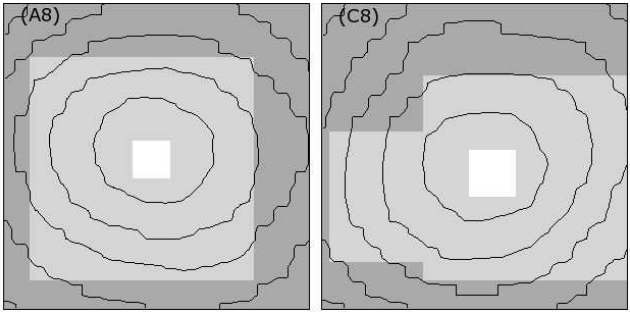}}
  \caption{Projections on the $xy$ plane, showing the projected AMR level in a zone with a side of $0.2\ R_\rmn{vir}$, centred at the cluster core, at $z = 0$. The colour white corresponds to the AMR level $l = 8$, light gray to $l = 7$ and dark gray to $l = 6$, respectively. Baryon density contours are superimposed. The two panels refer to the runs with maximum AMR level $l_\rmn{max} = 8$ and the AMR setup indicated by the letter at the upper left corners.} 
  \label{convergence-core}
\end{figure}

A more significant convergence test has been carried out by repeating the runs $A$ and $C$ with the same parameters and setup, but allowing (from $z = 2$) different maximum AMR levels. The grid coverage in the cluster core at $z = 0$ is shown in Fig.~\ref{convergence-core} for the runs with maximum AMR level $l_\rmn{max} = 8$, whereas the two runs with $l_\rmn{max} = 6$ have a complete grid coverage in the selected region. 

\begin{figure}
  \resizebox{\hsize}{!}{\includegraphics{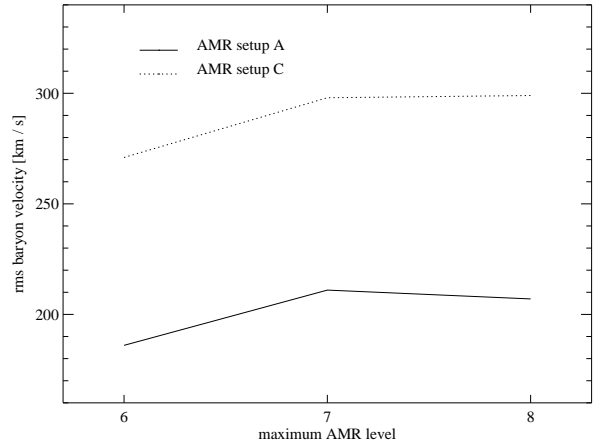}}
  \caption{Mass-weighted, rms baryon velocity calculated within a sphere with $R = 0.1\ R_{\rmn{vir}}$ centred at the cluster centre at $z = 0$, for runs with different AMR maximum levels (indicated on the $y$ axis) and refinement criteria (cf.~Table \ref{run}). Solid line: AMR criteria as in run $A$. Dotted line: AMR criteria as in run $C$.} 
  \label{convergence-rms}
\end{figure}

The rms velocities within the innermost $0.1\ R_{\rmn{vir}}$ are presented in Fig.~\ref{convergence-rms}. For each AMR setup, one can see that $v_{\rmn{rms}}(l_\rmn{max} = 6) < v_{\rmn{rms}}(l_\rmn{max} = 7)$, because the flow structures are less resolved by the coarser grid structure. On the other hand, the runs with $l_\rmn{max} = 8$ have a level of rms velocity which is the same as the simulations with $l_\rmn{max} = 7$. This is also due to the AMR grid, because the volume coverage of $l =8$ is very limited with the employed refinement criteria (Fig.~\ref{convergence-core}). 

It appears therefore that the level of velocity fluctuations in the cluster core, when averaged on a sphere of $R = 128\ \rmn{kpc}\ h^{-1}$, is not affected by setting the effective resolution below $7.8\ \rmn{kpc}\ h^{-1}$ (AMR level $l = 7$). The values of $v_{\rmn{rms}}$ show a clear convergence for increasing $l_\rmn{max}$. It is interesting to point out that $v_{\rmn{rms}}(A) < v_{\rmn{rms}}(C)$ even at $l_\rmn{max} = 6$, when the grid coverage of the cluster core is identical for the two AMR setups. From this convergence study we can assert that the effectiveness of the new AMR criteria in resolving turbulent flows does not come trivially from a larger grid coverage, or a larger number of grids, but instead from a more suitable refinement strategy. 

Though our simulations show convergence between the AMR levels 7 and 8, one should note that the number of DM particles (and thus the mass of each particle) could affect the results. This issue has not been fully explored in this work, but in a test runs with coarser particle resolution ($64^3$ DM particles at $l = 0$ and $1$) and AMR setup $A$, we found a velocity dispersion only $5\ \%$ smaller than in run $A$ with $l_\rmn{max} = 6$, which formally has the same effective spatial resolution.

\section{Summary and conclusions}
\label{conclusions}

The study of turbulent flows in the ICM is important for a
thorough understanding of galaxy clusters and for the physics of the
plasma in these objects. The turbulent state of the ICM is, from an
observational point of view, still a debated issue, and the
theoretical determination of the kinematic viscosity contains large
uncertainties. Besides these open problems, a specific question has
been addressed in this work: are the existing numerical techniques of
grid-based, AMR cosmological simulations suitable for studying the
main properties of turbulence in the ICM? In particular, our attention
was focused on the AMR criteria employed in simulations of
strongly clumped objects. A galaxy cluster was therefore simulated
with a standard, reference setup, and then the simulation was repeated
with five different choices of refinement criteria.  

The runs were subdivided into two groups, with an underlying difference in
the refinement strategy. In the first group, we used AMR criteria based
on the regional variability of control variables of the flow, designed
for refining the grid at the locations of significant velocity
fluctuations. In the second group, a more accurate tracking of the
overdense subclumps was enforced.  

The distinction between the two groups of runs is important for the
interpretation of the results of Sects.~\ref{filling} and \ref{core}. 
In the runs $B$, $C$ and $D$, the values of $\max{(|\bomega|)}$ (Table
\ref{vorticity}) and  $v_{\rmn{rms}}$ (Table \ref{core-table}) are
larger with respect to run $A$, and to the runs of the second
group. The AMR criteria used in these runs are therefore very suitable
for refining the grid where the velocity fluctuations are
underresolved, and result in an increased magnitude of the
turbulence. In the second group of runs, the AMR does not refine
explicitly on turbulence, but the lower threshold on overdense region
allows a better tracking of subclumps. In other words, the stirring
mechanism for turbulence generation in the ICM is better
resolved. Therefore comparatively larger $\langle |\bomega| \rangle$
and a larger volume-filling factor are obtained.  

In both groups of simulations, the results in the innermost $0.1\
R_{\rmn{vir}}$ are consistent with a better modelling of the turbulent
flow. In addition to an increased rms velocity, the change in
$P_{\rmn{turb}}/P_{\rmn{tot}}$ indicates that the turbulent component
plays a non-negligible role in the pressure support. Recently \citet{cfv07}, based on X-ray and optical observations of M87 and NGC 1399, put a constraint of the order of $10\ \%$ on the non-thermal contribution to the thermal pressure, due to turbulent motions, cosmic rays and magnetic fields. This is not in contradiction to the findings of our work, because we find that the turbulent pressure is only a few percent of the thermal pressure at most. 
%One of the usual methods for estimating the cluster mass from X-ray observations \citep{edm02} makes use of the assumption of hydrostatic equilibrium in spherical symmetry. According to the presented results, the importance of the turbulent pressure needs to be properly taken into account in order to avoid mass underestimations \citep{rem06,mhb07}. 

The presented comparison shows clearly that run $A$ (which implements
AMR criteria widely used in grid-based cosmological simulations) fails
to reproduce both the magnitude of the rms velocity and the spatial
extent of the turbulent flow. This central issue should be carefully
taken into account in future investigations of the ICM turbulent
features. These results suggest that computationally more
demanding simulations will be able to resolve more flow and
substructures, which in turn can further contribute to the turbulent
properties. 

Apparently the presented AMR approach to the modelling of turbulent
flows in the ICM has limited utility outside of the cluster core. As
discussed in Sec.~\ref{core}, this is partly due to the
volume-filling properties of turbulence in clusters, and partly to the
constraint on the number of produced AMR grids which, if too many,
would make the simulation computationally unaffordable.  

The presented results are similar to previous simulations
\citep{nb99,dvb05}, as far as the magnitude of the turbulent velocity
is concerned. Interestingly, \citet{k07} shows that the level of
ICM turbulence driven by galaxy motions saturates around 220 km/s,
which is not very different from our results.
As stated in the Introduction, a detailed comparison of
the features of turbulence in SPH and grid-based simulation is out of
the scope of this paper, but we observe that the trends that we
inferred from Table \ref{core-table} are also present in
\citet{dvb05}. In that case, when the low-viscosity SPH scheme is
used, the variation of some of the variables is more marked that in
our study. However it should be stressed that in this work we do not
make use of a improved numerical scheme to better resolve turbulence,
but only of more suitable choices of AMR criteria.  

In many astrophysical problems (including galaxy clusters) the range
of length scales needed to follow the turbulent cascade down to the
Kolmogorov dissipation scale extends well beyond the grid spatial
resolution limit, even using AMR. A strong improvement in the
modelling of turbulence would be the application of Large Eddy Simulations,
where only the largest scales are resolved and the dynamics at length
scales smaller than the spatial resolution is handled by a subgrid
scale model \citep{sag01,snh06,snhr06}. Such a tool will allow to
consistently evaluate the level of subgrid turbulent energy and will
make the numerical information about the level of turbulence in the
flow more directly available. We expect that the magnitude of the
turbulent motions, also outside $0.1\  R_{\rmn{vir}}$, could be
substantially increased.  
%This method obviously bypasses the issue of numerical viscosity,
%unavoidably introduced by numerically solving the Euler equations,
%but makes the comparison with the theoretical prediction for $\nu$
%more delicate.  
The use of this tool in the framework of cosmological simulations will
be explored in future work.

\section*{acknowledgements}

The numerical simulations were carried out on the SGI Altix 4700 {\it
 HLRB2} of the Leibnitz Computing Centre in Munich (Germany). Thanks
go to C.~Federrath and M.~Hupp for assistance with the analysis tools,
 to F.~Miniati and W.~Schmidt for interesting discussions, and to K.~Dolag for reading and commenting the manuscript. Our research was
supported by the Alfried Krupp Prize for Young 
University Teachers of the Alfried Krupp von Bohlen und Halbach
Foundation.

\bibliography{cluster-index}
\bibliographystyle{bibtex/mn-web}

\bsp

\label{lastpage}

\end{document}